\documentclass[aps,showpacs]{revtex4}
\usepackage{bm}
\usepackage{graphicx}
\usepackage{subfigure}

\begin{document}

\newcommand{\kb}{k_{\mathrm{B}}}
\newcommand{\etak}{\eta_{\mathrm{K}}}
\newcommand{\Rey}{\mathrm{Re}}
\newcommand{\Wi}{\mathrm{Wi}}
\newcommand{\St}{\mathrm{St}}
\newcommand{\taust}{\tau_\mathrm{st}}
\newcommand{\rhop}{\rho_{\mathrm{p}}}
\newcommand{\rhof}{\rho_{\mathrm{f}}}
\newcommand{\epsin}{\varepsilon_{\mathrm{in}}}
\newcommand{\epsp}{\varepsilon_{\mathrm{p}}}
\newcommand{\mb}{m_{\mathrm{b}}}
\newcommand{\np}{N_{\mathrm{p}}}
\newcommand{\knp}{n_{\mathrm{p}}}
\newcommand{\nup}{\nu_{\mathrm{p}}}
\newcommand{\etap}{\eta_{\mathrm{p}}}
\newcommand{\vb}{V_{\mathrm{b}}}
\newcommand{\fp}{{\bm f}_\mathrm{p}}

\title{Two-way coupling of FENE dumbbells with a turbulent shear flow}
\author{Thomas Peters\footnote{Present address: Institute for Theoretical Astrophysics, 
Ruprecht-Karls-Universit\"at Heidelberg, D-69120 Heidelberg, Germany}}
\affiliation{Department of Physics,
             Philipps-Universit\"at Marburg, D-35032 Marburg, Germany}
\author{J\"org Schumacher\footnote{Corresponding author: joerg.schumacher@tu-ilmenau.de}}
\affiliation{Department of Mechanical Engineering,
           Technische Universit\"at Ilmenau, D-98684 Ilmenau, Germany}
\date{\today}

\begin{abstract}
We present numerical studies for finitely extensible nonlinear elastic (FENE) dumbbells
which are dispersed in a turbulent plane shear flow at moderate Reynolds number. The
polymer ensemble is described on the mesoscopic level by a set of stochastic ordinary 
differential equations with Brownian noise. The dynamics of the Newtonian solvent is 
determined by the Navier-Stokes equations.
Momentum transfer of the dumbbells with the solvent is implemented by an additional volume
forcing term in the Navier-Stokes equations, such that both components of the resulting viscoelastic
fluid are connected by a two-way coupling. The dynamics of the dumbbells is given then by 
Newton's second law of motion including small inertia effects. We investigate
the dynamics of the flow for different degrees of dumbbell elasticity and inertia, 
as given by Weissenberg and Stokes numbers, respectively. For the parameters accessible
in our study, the magnitude of the feedback of the polymers on the macroscopic properties
of turbulence remains small as quantified by the global energy budget and the Reynolds
stresses. A reduction of the turbulent drag by up to 20$\%$ is observed for the larger
particle inertia. The angular statistics of the dumbbells shows an increasing alignment
with the mean flow direction for both, increasing elasticity and inertia. This goes in
line with a growing asymmetry of the probability density function of the transverse derivative of
the streamwise turbulent velocity component. We find that dumbbells get stretched 
preferentially in regions where vortex stretching or bi-axial strain dominate the local
dynamics and topology of the velocity gradient tensor.
  
\noindent 
\end{abstract}

\maketitle

\section{Introduction}
When a few parts per million in weight of long-chained polymers are added to a 
turbulent fluid its properties change drastically and a significant reduction 
of turbulent drag is observed. \cite{Lumley1969} Although the phenomenon is known 
from pipe flow experiments for almost 60 years,\cite{Toms1949,Virk1975} a complete 
understanding is still lacking. One reason for this circumstance is that the physical
processes in a turbulent and dilute polymer solution cover several orders of magnitude in 
space and time; in other words, we are faced with a real multiscale problem. 
\cite{McKinley2002,Larson2005} In case of fully developed turbulence, the integral scale 
$L$, which measures the extension of largest vortex structures in the flow, exceeds the 
viscous Kolmogorov scale $\etak$, which stands for the extension of the smallest turbulent
eddies,  by a factor of at least 1000. However, long-chained polymers barely exceed 
the viscous flow scale even in an almost stretched state. Their equilibrium extension as 
given by the Flory radius $R_0$ is usually by a factor of 100 smaller than 
$\etak$.\cite{Doi1996} In terms of time scales the situation differs slightly. The viscous
Kolmogorov time $\tau_{\eta}$ can become smaller than the slowest relaxation time $\tau$ of the
macromolecules. Although macroscopic closures can rationalize some issues of drag
reduction \cite{Benzi2006}, the challenging question remains of how the individual dynamics
of numerous polymer chains, which is present on sub-Kolmogorov and Kolmogorov scales, adds up 
to a macroscopic effect at scales $r\lesssim L$ as being observed in several experiments. 
\cite{Warholic1999,White2000,White2004}

The description of dilute polymer solutions relies for most studies on one of the following 
two models: on one side, {\em macroscopic continuum models} such as Oldroyd-B or FENE-P models 
\cite{Bird1987,Sureshkumar1997,Ilg2002,Eckhardt2002,Dimitropolous2005} include the
polymer dynamics as an additional additive macroscopic stress field. Only the largest scales 
$\ell \gtrsim \etak$ of the viscoelastic fluid are described in its full complexity.
Numerical problems arise in connection with the pure hyperbolic character of the equation of
motion for the polymer stress field, such as the conservation of its positivity (see e.g.
Ref.~\cite{Vaithianathan2003} for a detailed discussion). 
In addition, the coarse graining to the macroscopic
polymer stress can lead to deeper conceptional difficulties, e.g., the failure of energy stability
of viscoelastic flows, which is an important building block for investigations of stability and upper
bounds on the dissipation rate in Newtonian flows. \cite{Doering2006}
Further problems arise for the macroscopic description of non-Newtonian fluids in the limits
of very low and high frequencies, where they should behave as Newtonian fluids and solids, 
respectively. \cite{Pleiner2000,Beris2001,Pleiner2001}

On the other side, {\it Brownian dynamics models} 
\cite{Oettinger1996,Puliafito2005,Hur01,Graham2003,Terrapon2004} describe the polymer chain on a
mesoscopic level as overdamped coupled oscillators arranged in bead-spring chains. The models
include complex conformations of the macromolecules and screening effects due to the solvent 
such as hydrodynamic interaction.\cite{Zimm1956} The simplest of such mesoscopic models 
for a polymer chain is a dumbbell where two beads are connected by a spring. The dynamics in
these models is on scales $\ell\lesssim \etak$. This means that the surrounding fluid is
spatially smooth and either a steady \cite{Puliafito2005}, a start-up shear flow \cite{Hur01},
or a white-in-time random flow. \cite{Celani2005} In a recent work by Davoudi and 
Schumacher\cite{Schumacher2006}, numerical studies at the interface of both descriptions were conducted by 
combining Brownian dynamics simulations (BDS) with direct numerical simulations of a turbulent 
Navier-Stokes shear flow. The simplest mesoscopic model with a linear spring force - the
Hookean dumbbell model - was taken there in order to study the stretching of the dumbbell as
a function of the outer shear rate and the elastic properties of the springs. However,
a feedback of the polymers on the shear flow was not included in their study.  
  
In the following, we want to extend these investigations into two directions. Firstly, we will 
model the macromolecules more realistically as finitely extensible nonlinear elastic (FENE) 
dumbbells. Secondly, their feedback on the shear flow is included via a two-way 
coupling. The effect of the FENE dumbbells on the statistical fluctuations of the velocity and 
the velocity gradients will be studied. In addition, conformational properties of the dumbbells, 
such as their extension and angular distribution with respect to the mean flow component, will be 
addressed.  The polymer feedback results in an additional forcing that 
has to be added to the right hand side 
of the Navier-Stokes equations for the advecting Newtonian solvent similar to the case of two-phase
flows with dispersed particles \cite{Squires1990,Elgobashi1993,Bosse2006} or
bubbles.\cite{Mazzitelli2003} We will keep the full dynamic equation of motion
for the dumbbells, containing accelerations due to elastic, friction and stochastic forces, 
and cannot neglect inertia. This step is necessary in order to describe the momentum transfer
of the dumbbells to the solvent as discussed in Ref.\cite{Ahlrichs1999}. 

In contrast to the conventional BDS that neglect inertia effects from beginning, we will be left 
here with three physical parameters: the Stokes number $\St$ for the particle inertia, the Weissenberg 
number $\Wi$ for the elastic properties of the dumbbells, and the Reynolds number $\Rey$ of the flow, 
respectively. The Reynolds number is defined as  
\begin{equation}
\Rey=\frac{U L}{\nu}\,,
\label{Rey}
\end{equation}
with the characteristic (large-scale) velocity $U$, the characteristic length $L$ (both are specified 
later in the text), and the kinematic viscosity of the Newtonian solvent $\nu$. The Weissenberg
number $\Wi$ compares the characteristic dumbbell relaxation time $\tau$ from a stretched
to a coiled state with the characteristic time scale of the advecting flow, $L/U$, and is given 
by 
\begin{equation}
\Wi=\frac{U \tau}{L}\,.
\label{Weiss}
\end{equation}
The Stokes number $\St$ relates the particle response time to changes in the surrounding velocity, 
$\taust$, with the characteristic flow time scale. It follows to
\begin{equation}
\St=\frac{U \taust}{L}\,.
\label{Stokes}
\end{equation}
The physics of dispersed FENE dumbbells in a turbulent shear flow is thus
described by three dimensionless numbers. For a fixed Reynolds numbers $\Rey$, we can basically
distinguish the following four limiting cases: (i) $\Wi\gg 1$, $\St\gg 1$; (ii) $\Wi\ll 1$,
$\St\gg 1$; (iii) $\Wi\ll 1$, $\St\ll 1$; (iv) $\Wi\gg 1$, $\St\ll 1$. Case (i) would stand for very
heavy particles (or dumbbells) which are stretched almost to their contour length. They will
behave as dispersed rods. In case (ii), the dumbbells would act as heavy spherical 
particles since they remain coiled in practical terms. The cases of interest for dilute polymer
solutions are (iii) and (iv), respectively. Inertia effects are then very small, \cite{Schieber1988} and
the Weissenberg number can vary from very small to large values implying an increasingly slower 
relaxation of the macromolecules from a stretched non-equilibrium to a coiled equilibrium state 
in comparison to the characteristic flow variation time scale. As we will discuss in the next 
section, the numerical treatment becomes challenging, on one hand due to the 
finite extensibility, on the other hand due to the small Stokes numbers we are aiming at. The Stokes
time $\taust$ sets a small but finite time scale then, which can cause stiffness problems for an
explicit integration algorithm. Despite these efforts, our values 
for the Stokes number will still exceed the realistic magnitudes for polymer chains in solution
by orders of magnitude. Nevertheless, we think it is interesting and to some degree necessary to 
study the dumbbell dynamics under these circumstances and to provide a systematic study of how a 
shear flow will be affected by the presence of dispersed bead-spring chains with variable degree
of inertia. This will shed some light on possible reasons for drag reduction in our model. 

The outline of the manuscript is as follows. In the next section the equations of motion,
the two-way coupling and the numerical scheme are presented. Afterwards, we discuss
the results for the macroscopic energy balance as well as for the Reynolds stresses. This is
followed by studies of small-scale properties such as the statistics of the extension and 
orientation of the dumbbells and of their impact on the fluctuations of velocity gradients. We 
conclude with a discussion of our results and will give a brief outlook to extensions of the 
present work toward more realistic parameter settings.

\section{Model and equations}
\subsection{The Newtonian solvent}
The Navier-Stokes equations that describe the dynamics of the three-dimensional incompressible
Newtonian fluid
are solved by a pseudo-spectral method using a second-order predictor-corrector 
scheme for advancement in time.\cite{Schumacher2006} The equations of motion are 
\begin{eqnarray}
\label{nseq}
\frac{\partial{\bm u}}{\partial t}+({\bm u}\cdot{\bm \nabla}){\bm u}
&=&-{\bm \nabla} p+\nu {\bm \nabla}^2{\bm u}+{\bm f} +\fp\,,\\
\label{ceq}
{\bm \nabla}\cdot{\bm u}&=&0\,,
\end{eqnarray}
where ${\bm u}$ is the (total) velocity field, $p$ the kinematic pressure field, ${\bm f}$
the volume forcing which sustains the turbulence, and $\fp$ the feedback of the dumbbells
(see section~II~C). The shear flow is 
modeled in a volume with free-slip boundary conditions in the shear direction $y$ and periodic 
boundaries in the streamwise and spanwise directions $x$ and $z$. The free-slip boundary conditions
at $y=0, L_y$ are given by
\begin{equation}
u_y=0\,,\;\;\;\qquad\frac{\partial u_x}{\partial y}=\frac{\partial u_z}{\partial y}=0\,.
\end{equation}
Here, the total velocity 
field follows by a Reynolds (de)composition as a linear mean part with the constant shear
rate $S$ and a turbulent fluctuating part
\begin{equation}
{\bm u}=\langle{\bm u}\rangle + {\bm u}^{\prime}=Sy{\bm e}_x+{\bm u}^{\prime}\,.
\label{Reynolds}
\end{equation}
The notation $\langle\cdot\rangle$ stands for the ensemble average, which will be a combination 
of volume and time averages for most cases. The aspect ratio is $L_x \colon L_y \colon L_z
=4\pi \colon 2 \colon 2\pi$.
The characteristic length is the halfwidth of the slab, $L=L_y/2$. Velocities are measured in 
units of the laminar flow profile ${\bm U}(y)=-\sqrt{2}\cos(\pi y/2){\bm e}_x$. We will take
$U_x(L_y/4)$ as the characteristic velocity $U$ (see also (\ref{Rey}), (\ref{Weiss}), and (\ref{Stokes})).
The applied volume forcing sustains this laminar flow profile and follows from (\ref{nseq})
consequently to ${\bm f}(y)=-\sqrt{2}\pi^2/(4\nu)\cos(\pi y/2){\bm e}_x$. Forcing amplitude and
profile will remain unchanged throughout this study. At sufficiently
large Reynolds numbers this linearly stable laminar shear flow becomes turbulent when a finite
perturbation is applied.\cite{Schumacher2001} The volume forcing $\bm f$ 
is then a permanent source of kinetic energy injection into the shear flow which sustains turbulence 
in a statistically stationary state. Although the steady forcing is of 
cosine shape, the resulting mean turbulent flow profile will be linear except for small layers 
in the vicinity of both free-slip planes, where the boundary conditions have to be satisfied.
Our mean profiles follow to $\langle u_x(y)\rangle \simeq S(y-1)$ for $y\in [0,2]$ 
with $S=0.035-0.04$ for $\Rey=800$. This range of $S$-values remained nearly unchanged
for all parameter sets. In addition, $\langle u_y^{\prime}\rangle=\langle u_z^{\prime}\rangle=0$.
The shear flow can be considered therefore as being {\em nearly} homogeneous.

The simulation program is run with two spectral resolutions. For $\Rey=400$, a grid with
$64\times 32 \times 32$ mesh points was taken. For $\Rey=800$, we took a grid with
$128\times 32\times 64$ points. The spectral resolution as given by the product 
$k_{\mathrm{max}}\etak=\sqrt{8}\pi N_x/(3 L_x)\etak$ was 1.5 for the first case and 2.3 for the
second. Here, $\etak$ is the viscous Kolmogorov scale and defined as
$\etak=\nu^{3/4}/\langle\varepsilon^{\prime}\rangle^{1/4}$ with the mean turbulent energy
dissipation rate $\langle\varepsilon^{\prime}\rangle$, where $\varepsilon^{\prime}({\bm x},t)=
(\nu/2)(\partial u^{\prime}_i/\partial x_j+\partial u^{\prime}_j/\partial x_i)^2$ for 
$i,j=x,y,z$. Clearly, the spectral resolutions are not very large, but they give us the 
opportunity to perform parametric studies in the three-dimensional space which is spanned
by $\Rey$, $\Wi$, and $\St$. Most of our following studies will be conducted for the better
resolved case of $\Rey=800$.

\subsection{The FENE dumbbells}
The smallest building block for the mesoscopic description of the polymer stretching can be 
accomplished by considering dumbbells where two beads (that stand for several hundreds of 
monomers) are connected by a spring. The entropic elastic force follows the Warner force 
law \cite{Bird1987} and depends on the separation vector ${\bm R}(t)={\bm x}_2(t)-{\bm x}_1(t)$
that is spanned between both beads at positions ${\bm x}_2(t)$ and ${\bm x}_1(t)$, respectively.
The force law is given by 
\begin{equation}
{\bm F}_\mathrm{el}({\bm R})=\frac{H {\bm R}}{1-R^2/L_0^2}\,,
\end{equation}
where $L_0$ is the contour length of the dumbbells which cannot be exceeded. 
The spring constant is denoted by $H$. When taking
into account the elastic entropic force, hydrodynamic Stokes drag, and thermal noise,
the second Newtonian law for a FENE dumbbell written in relative coordinates ${\bm R}(t)$
and center-of-mass coordinates ${\bm r}(t)=({\bm x}_1(t)+{\bm x}_2(t))/2$ reads
\cite{Schieber1988,Celani2005}
\begin{eqnarray}
\dot{\bm r}&=&{\bm v}\,, \label{r1}\\
\frac{\mb}{\zeta} \dot{\bm v}&=&-{\bm v}+\frac{1}{2}({\bm u}_1+{\bm u}_2)+
\sqrt{\frac{\kb T}{\zeta}}{\bm \xi}_{\bm r}\,, \label{r2}\\
\dot{\bm R}&=&{\bm V}\,, \label{r3}\\
\frac{\mb}{\zeta} \dot{\bm V}&=&-{\bm V}+\Delta{\bm u}-\frac{2H{\bm R}}
{\zeta\left(1-R^2/L_0^2\right)}+
\sqrt{\frac{4 \kb T}{\zeta}}{\bm \xi}_{\bm R}\,,\label{r4}
\end{eqnarray}
where $\Delta {\bm u}= {\bm u}({\bm x}_2,t)-{\bm u}({\bm x}_1,t)$ is the relative fluid velocity
at the bead centers. The last terms in the velocity equations, containing ${\bm \xi}_{\bm r}$ and 
${\bm \xi}_{\bm R}$, stand for vectors of thermal Gaussian noise with the properties
\begin{eqnarray}
\langle\xi_i(t)\rangle&=&0\,,\\
\langle\xi_i(t)\xi_j(t^{\prime}\rangle&=&\delta_{ij}\delta(t-t^{\prime})\,
\label{white}
\end{eqnarray}
for $i,j=x,y,z$. The three components of each vectorial noise term are statistically independent 
stochastic processes. Furthermore, the vectorial noise with respect to the center-of-mass velocity 
is statistically independent to that for the relative velocity dynamics. 
The noise prevents the extension of a dumbbell to shrink below its equilibrium length 
\begin{eqnarray}
R_0=\sqrt{\frac{\kb T}{H}}\,,
\label{equilibrium}
\end{eqnarray}
with $\kb$ being the Boltzmann constant, $T$ the temperature. Equation (\ref{equilibrium})
follows from the equipartition theorem. The contour length $L_0=10 R_0$ is used throughout this
study and $R_0\simeq \etak$.  The relaxation time of the dumbbells is given by 
\cite{Bird1987}
\begin{equation}
\tau=\frac{\zeta}{4H},
\label{relaxationtime}
\end{equation}
where 
\begin{equation}
\zeta=6\pi\rhof\nu a
\label{frictioncoefficient}
\end{equation}
is the Stokes drag coefficient of a spherical bead with radius
$a$. The fluid mass density is $\rhof$. Due to the current resolution contraints the dumbbells
will experience both the smooth and partly rough scales of the advecting flow. Consequently, 
the velocity difference $\Delta {\bm u}$ is kept in the equation and not approximated by 
the linearization $\Delta {\bm u}\approx({\bm R\cdot \nabla}){\bm u}$ as it is done in BDS where 
$L_0\ll \eta_K$. For spatially smooth flows both expressions give the same results.

The equations (\ref{r1}) through (\ref{r4}) introduce the other two dimensionless parameters
beside the Reynolds number $\Rey$, the Weissenberg number $\Wi$ and the Stokes number $\St$, 
respectively (see definitions (\ref{Weiss}) and (\ref{Stokes})). 
The Stokes time $\taust$ is the response time of an inertial particle which is
required to speed up to the velocity of its local surrounding. A zero Stokes time
implies a behavior as a passive Lagrangian tracer. For beads, this time follows to 
$\taust=\mb/\zeta$ with $\zeta$ as given above and consequently
\begin{equation}
\taust=\frac{2 \rhop a^2}{9 \rhof \nu}\,.
\end{equation}
The density contrast $\rhop/\rhof$ is to very good approximation unity\cite{Flory},
i.e. polymers are considered as neutrally buoyant. 
In Ref.~\cite{Schumacher2006}, we have compared the polymer 
relaxation time to the microscopic stretching time
scale. This is given by the inverse of the maximum Lyapunov exponent and is comparable
to the microscopic time scale of the flow, the Kolmogorov time $\tau_{\eta}=\sqrt{
\nu/\langle\varepsilon\rangle}$. Table 1 gives an overview of the values of $\St$ and
$\Wi$ that have been used and of how they translate into $\St_{\eta}$ and $\Wi_{\eta}$,
respectively. We see that the Stokes numbers get as low as $10^{-4}$ when measured in 
viscous units, which is still orders of magnitude above the realistic estimates for
dilute polymer solutions which are about three to four order of magnitude below our minimal
value.  

In most cases, an ensemble of $6.3\times 10^4$ FENE dumbbells, i.e. $1.2\times 10^5$ beads, is
advanced by a weak second-order predictor-corrector scheme simultaneously with the flow 
equations.\cite{Oettinger1996}
The finite extensibility and the small Stokes numbers require a semi-implicit time-stepping
for some variables. In order to avoid a total length larger than $L_0$, we proceed
in line with Ref.~\cite{Oettinger1996} and solve a cubic
equation for $R=|{\bm R}|$ in the corrector step. 
Initially, the center of mass of the dumbbells is seeded randomly in space with a 
uniform distribution and an initial extension of $R_0$. All Lagrangian interpolations were done
with a trilinear scheme. Details on the numerical procedure are outlined in appendix A.

In order to build a bridge to macroscopic simulations we provide an estimate for the contribution of the
dumbbell ensemble to the zero-shear viscosity. Following Ref.~\cite{Oettinger1996} it is defined
as 
\begin{equation}
\etap=\rhop\nup=\knp \kb T \tau\,,
\end{equation}
with the number density of dumbbells $\knp$. When applying (\ref{equilibrium}) as well as 
definitions (\ref{relaxationtime}) and (\ref{frictioncoefficient}), and using $\rhof/\rhop=1$ one gets
\begin{equation}
\nup=\frac{3}{2}\pi \knp R_0^2 \nu a\,
\end{equation}
with the solvent viscosity $\nu$.
The bead radius $a$ is substituted
by the Stokes time $\taust$. Recalling the definitions 
for the Kolmogorov length $\etak=\nu^{3/4}/\langle\varepsilon^{\prime}\rangle^{1/4}$
and for the Kolmogorov time $\tau_{\eta}=\sqrt{\nu/\langle\varepsilon^{\prime}\rangle}$,
one ends with the relative viscosity
\begin{equation}
s=\frac{\nup}{\nu}=\frac{9\pi}{2\sqrt{2}} \knp R_0^2 \etak \sqrt{\St_{\eta}}\,.
\label{ratio}
\end{equation}
For the present simulations, one dumbbell is seeded per grid cell and 
therefore $\knp\approx 1/\etak^3$. Additionally, $R_0\simeq\etak$. 
Following table 1 for the runs at $\Rey=800$, one gets ratios of $s$ between
between 0.1 for the smallest Stokes number going up to 3 for the largest one. The latter
value is rather large for polymer solutions. Values below unity
are usually taken, such as in DNS with the Oldroyd-B model.\cite{Eckhardt2002}
Equation (\ref{ratio}) is in this spirit consistent with the discussion in the introductory part.
Only the lower Stokes numbers result to values of $s$ as taken for macroscopic DNS
for viscoelastic shear flows. 
\renewcommand{\arraystretch}{1.5}
\begin{table}
\begin{center}
\begin{tabular}{lll}
\hline\hline
           & $\Rey=400$ & $\Rey=800$\\
\hline
$\Wi=3$   & $\Wi_{\eta} = 0.8$ & $\Wi_{\eta} = 0.6$ \\
$\Wi=20$  & $\Wi_{\eta} = 5.1$ & $\Wi_{\eta} = 4.3$ \\
$\Wi=100$ & $\Wi_{\eta} = 25.7$ & $\Wi_{\eta} = 21.5$ \\
\hline
$\St=5.0\times10^{-4}$ & $\St_{\eta}=1.3\times 10^{-4}$ & $\St_{\eta}=1.1\times 10^{-4}$ \\
$\St=5.0\times10^{-3}$ & $\St_{\eta}=1.3\times 10^{-3}$ & $\St_{\eta}=1.1\times 10^{-3}$ \\
$\St=5.0\times10^{-2}$ & $\St_{\eta}=1.3\times 10^{-2}$ & $\St_{\eta}=1.1\times 10^{-2}$ \\
$\St=5.0\times10^{-1}$ & $\St_{\eta}=1.3\times 10^{-1}$ & $\St_{\eta}=1.1\times 10^{-1}$ \\
\hline\hline
\end{tabular}
\caption{The Weissenberg and Stokes numbers rescaled by the Kolmogorov time $\tau_{\eta}$ of the flow.
$\Wi_{\eta} = \tau /\tau_{\eta}$ and $\St_{\eta}=\taust/\tau_{\eta}$. Note that $\tau_{\eta}$ is
based on the pure Newtonian case. Only minor changes arise when polymers are added to the solvent.
}
\end{center}
\label{table1}
\end{table}

\subsection{Two-way coupling}
The back-reaction of the dumbbells on the fluid consists of contributions from the Stokes
friction and the stochastic noise term. In accordance with Newton's third law, the force 
contribution from each of the two beads at positions ${\bm x}_i$ ($i=1,2$) follows to
\begin{equation}
{\bm F}_i=-{\bm F}_i^{(st)}-{\bm F}_i^{(n)}
         =\zeta(\dot{\bm x}_i-{\bm u}({\bm x}_i))-\sqrt{2\kb T\zeta}\,{\bm \xi}_i\,.
\end{equation}
The force density generated by all FENE dumbbells results to
\begin{equation}
\rhof \fp=\sum_{j=1}^{\np}\sum_{i=1}^2 {\bm F}_i^{(j)}\delta({\bm x}-{\bm x}_i^{(j)})\,,
\label{feedback1}
\end{equation}
where $\np$ is the number of dumbbells. The volume integral of (\ref{feedback1}) gives a force
since the delta function carries the dimension of an inverse volume due to $\int\delta({\bm x}-
{\bm x}_i^{(j)})\,\mbox{d}^3x=1$. Consequently, the dimensionless form of the forcing reads
\begin{equation}
\fp=\frac{\vb}{L^3 \St}\sum_{j=1}^{\np}\sum_{i=1}^2
\left[(\dot{\bm x}_i^{(j)}-{\bm u}({\bm x}_i^{(j)}))-\frac{R_0}{\sqrt{\Wi}\,L}\xi_i^{(j)}
\right]\,
\tilde{\delta}({\bm x}-{\bm x}_i^{(j)})\,,
\label{feedback2}
\end{equation}
where the bead volume follows to $\vb=4\pi a^3/3=(4\pi/3)(9\nu\taust/2)^{3/2}$.
The notation $\tilde{\delta}$ is for the dimensionless delta function. We have used again 
$\rhof/\rhop\approx 1$.
The force density has to be evaluated at space points that are between the mesh vertices. Again
the trilinear interpolation has to be used to evaluate the contributions of the point force to 
the eight next neighboring mesh vertices.  
  
\section{Large-scale properties}
\subsection{Energy balance}
The first analysis step is the study of the effects of the two-way coupling on the macroscopic
properties of turbulence. Given the boundary conditions for our problem, eq.~(\ref{nseq})
results in the following balance for the total kinetic energy $E(t)=\frac{1}{2V}\int_V |{\bm u}|^2\,
\mbox{d}^3x$ with $V=L_x L_y L_z$,
\begin{eqnarray}
\frac{d E}{d t}&=&-\nu\langle(\partial u_i/\partial x_j)^2\rangle_V
+\langle {\bm u\cdot \bm f}\rangle_V+
\langle {\bm u\cdot \fp}\rangle_V\,,\nonumber\\
               &=&-\varepsilon(t)+ \epsin(t)-\epsp(t)
\label{energy}
\end{eqnarray}
where $\langle\cdot\rangle_V=\frac{1}{V}\int\cdot\,\mbox{d}^3x$ is the short notation for the
volume average. In case of statistical stationarity, one gets $\mbox{d}\langle E\rangle_t/
\mbox{d}t=0$ and thus
\begin{equation}
\langle\epsin\rangle=\langle\varepsilon\rangle+\langle\epsp\rangle\,.
\label{energy1}
\end{equation}
Figure~\ref{fig1} shows the three mean rates as a function of the Stokes number for two
Weissenberg numbers $\Wi=20, 100$. The mean energy dissipation rate $\langle\varepsilon\rangle$
and the mean energy injection rate $\langle\epsin\rangle$ are of the
same order of magnitude for all cases. They remain nearly unchanged with respect to Weissenberg number,
which indicates that the effect of the dumbbell ensemble on the macroscopic flow properties 
is small. Nevertheless, one observes a slight increase of the mean energy injection rate 
$\langle\epsin\rangle$ with respect to $\St$ going in line with a decrease of
$\langle\varepsilon\rangle$ (see upper and mid panel of Fig.~\ref{fig1}). Recall that the
energy injection rate will be maximal for the laminar case, i.e. for ${\bm u}\parallel{\bm f}$.
The trend of the data indicates that the streamwise flow component relaminarizes slightly 
with growing inertia. The lower panel of the same figure shows the findings for the dissipation 
due to polymer stretching $\langle\epsp\rangle$. As an additional energy dissipation
mechanism, it consumes injected energy which goes into the elastic energy budget of the 
dumbbell ensemble. The rate $\langle\epsp\rangle$ grows in magnitude with respect to both
parameters, the Stokes and Weissenberg number. For $\Wi=3$, the dumbbells are not significantly
extended and no clear trend of $\langle\epsp\rangle$ with $\St$ could be observed. The
dissipation rate $\langle\epsp\rangle$ is significantly smaller in comparison to the 
runs with larger $\Wi$.

In order to estimate the maximum feedback of the dumbbells on the flow, we
performed an ``academic experiment" for our system by tethering one of the two beads of a dumbbell at 
a fixed position. The dumbbells get then stretched more efficiently and undergo strong 
conformational fluctuations. Figure 2 illustrates their dramatic effect on the total kinetic 
energy. We compare the freely draining case with the tethered one and observe a significant 
decrease of the kinetic energy. An inspection of the flow structures indicates that the 
turbulent fluctuations are supressed almost completely. The flow becomes basically
laminar. The magnitude of the feedback for freely draining dumbbells will always remain 
significantly below this artifical limit with tethered dumbbells.

\subsection{Reynolds stresses}
Figure~\ref{fig3} shows the four non-vanishing components of the Reynolds stress tensor
$\langle u_i^{\prime} u_j^{\prime}\rangle/(2k)$ where $k=\langle (u_i^{\prime})^2\rangle/2$
is the turbulent kinetic energy (TKE). The moments are averages over the whole simulation 
volume for a sequence of about 100 statistically independent snapshots of the time evolution
of the shear flow. The results can be summarized to the following trends. 
For the two smallest Stokes numbers, no dependence on the Weissenberg number is observed.
For $\St=0.05$ and 0.5, the mean streamwise fluctuations are enhanced while the remaining
components of the Reynolds stress tensor decrease as a function of $\Wi$. This finding is in
agreement with observations in a Kolmogorov flow by Boffetta {\it et al.} \cite{Boffetta2005}

Similar to the friction factor for a turbulent pipe \cite{Schlichting}, we can define a 
friction factor for the present flow where the applied pressure gradient term has to be 
substituted by an amplitude of the static volume forcing profile ${\bm f}$ that sustains
the laminar cosine flow profile. Consequently,
\begin{eqnarray}
c_f=\frac{2 F L_y}{\langle u_x(y=L_y)\rangle^2}\,.
\label{friction}
\end{eqnarray}
Since ${\bm f}(y)=-\sqrt{2}\pi^2/(4\nu)\cos(\pi y/2){\bm e}_x$, we take $F=f_x(y=L_y)=
\sqrt{2}\pi^2/(4\nu)$. A similar definition was suggested for a Kolmogorov flow which is also
driven by a volume forcing.\cite{Boffetta2005} Drag reduction by dispersed dumbbells would go 
in line with a decrease of the dimensionless measure $c_f$ below the Newtonian value $c_f^N$. 
For the smallest Stokes number, the ratio goes to about unity. The slight overshoot is
attributed to the strong variations of the streamwise velocity at the free-slip planes.
Figure~\ref{fig4} indicates a reduction by $20\%-25\%$ at $\St=0.05, 0.5$ and for the larger
Weissenberg numbers. The series with $\Wi=3$ gave $c_f\simeq c_f^N$.

An important structural ingredient of shear flows are the asymmetric fluctuations of the three
diagonal elements of the Reynolds stress tensor. The streamwise fluctuations $\langle (u_x^
{\prime})^2\rangle$ are spatially arranged in streamwise streaks which interact with streamwise
vortices in a so-called regeneration cycle of coherent structures. This 
cycle is sustained by the non-normal amplification mechanism.\cite{Waleffe1997,Grossmann2001} 
The impact of long-chained polymers on the extension of the streamwise streaks has been demonstrated 
in experiments \cite{White2004} and numerical simulations.\cite{Stone2004,Dubief2004} While
streamwise fluctuations were found to increase, the fluctuations in shear and spanwise directions 
decreased. 
This is in line with our observations as discussed above. In Fig.~\ref{fig5}, we show
isolevels of the streamwise turbulent fluctuations for opposite sign at $\Wi=3, 20, 100$. Although not 
very pronounced, a slight increase 
in the connectivity and extension of the streamwise streaks can be observed with increasing 
Weissenberg number. 

As we can see, the statistics of macroscopic turbulent properties is affected only slightly by the
dispersed FENE-dumbbells. Their impact increases with Weissenberg number as well as with Stokes number.
In order to rule out that particle inertia dominates the discussed trends of our studies, 
we considered the case of dispersed beads in the same flow at the same Stokes numbers. This is achieved
by switching off the elastic
spring force, i.e. $F_\mathrm{el}=0$. The Stokes friction force remained as the only force.
The quantity $\fp$ models then 
the feedback of the particles on the flow. We added the statistical means of injection and 
dissipation rates as a function of the Stokes number for this case to Fig.~1. While the mean injection
and mean dissipation rates are of the same magnitude, the dissipation due to particle feedback is orders
of magnitude smaller in comparison to the polymer feedback, except for the largest $\St$. In addition,
we found no clear trends for the Reynolds stress components as a function of $\St$.

\section{Small-scale properties}
\subsection{Extensional and angular statistics of dumbbells}
The finite extensibility of the dumbbells will affect the shape of the probability density function 
(PDF) of $R$, which is supported on scales smaller than $L_0$ only. Figure~\ref{fig6} reports our
findings for $p(R)$ for different Weissenberg and Stokes numbers. For the lowest Weissenberg number,
$\Wi=3$, the majority of the dumbbells remains at the extension of about the Kolmogorov length $\etak$.
This picture changes for larger values of $\Wi$. At $\Wi=100$, the majority of the ensemble is stretched
to almost $L_0$, which manifests in the sharp maximum at $R\lesssim L_0$. Qualitatively, the change
of the shapes of the PDFs with increasing $\Wi$ agrees well with experimental findings \cite{Steinberg2005}
and analytical studies \cite{Celani2005,Chertkov2005} for the coil-stretch transition in random flows. 
The trends with the Stokes number remain small in all cases. However, the data show that growing
particle inertia suppresses the stretching to very extended molecules since the response time of the
molecules to the variation of the structures increases (see e.g. mid panel of Fig.~\ref{fig6}).

As we have seen in the last section, the fluctuations of the turbulent velocity field in the shear 
flow vary strongly from one space direction to another (see e.g. Fig.~\ref{fig3}). The major contribution
is contained in the streamwise component $\langle(u_x^{\prime})^2\rangle$ parallel to the 
direction of the mean turbulent flow. This suggests an 
investigation of the angular statistics of the polymers since their stretching can be expected to 
become anisotropic as well. The following dumbbell coordinate system will be used therefore throughout
this text: $R_x=R\cos\varphi\cos\theta$, $R_y=R\sin\varphi\cos\theta$, and $R_z=R\sin\theta$, where $R$ is
the distance between both beads. The notation differs from conventional spherical coordinates, but 
has the advantage of giving perfect alignment with the outer mean flow direction for $\varphi=\theta=0$.
$\varphi$ is the {\it azimuthal} angle and $\theta$ the {\it polar} angle. While the azimuthal angle
always remains in the shear plane that is spanned by the streamwise and shear directions, the polar
angle $\theta\ne 0$ indicates a dumbbell orientation out of this plane.

Davoudi and Schumacher \cite{Schumacher2006} discussed the statistics of both angles as a function 
of the Weissenberg number for passively advected Hookean dumbbells. The PDF of the polar angle was 
found to remain symmetric and to be less sensitive with respect to variations of $\Wi$. Our focus
will be therefore on the statistics of the azimuthal angle $\varphi$ which can take values between
$-\pi/2$ and $\pi/2$. The asymmetry between both quadrants is quantified by the following measure 
for the PDF $p(\varphi)$:
\begin{equation}
A(\varphi)=p(\varphi)-p(-\varphi)\,,
\label{asymmetry}
\end{equation}
with $\varphi\in[0,\pi/2]$. The measure $A(\varphi)$ is plotted for two Weissenberg numbers
in Fig.~\ref{fig7}. A pronounced maximum of $A(\varphi)$ implies that the dumbbells are
preferentially slightly tilted in the direction of shear, away from the mean flow direction (see
an illustration in Fig.~\ref{fig8}). We find that with 
increasing Weissenberg number the asymmetry of the angular
distribution grows in magnitude. The same trend holds when the Stokes number grows at fixed Weissenberg
number. In each case, the graph of $A(\varphi)$ shows an increasingly sharper maximum, which is shifted
towards smaller $\varphi$. Fluctuations of the dumbbells in the vicinity of $\varphi=0$ are enhanced 
while the tails for very large $\varphi$ are depleted. Growing inertia amplifies this trend. Once the 
dumbbells are aligned along the mean flow they remain in this orientation for longer periods of 
their evolution.

\subsection{Velocity gradient statistics}
Since the polymer dynamics takes place at the smallest scales of the turbulent flow, 
we study the impact of the dumbbells on the small-scale statistical 
properties of the flow in the following. 
Recent experimental and numerical studies in simple Newtonian shear flows indicate that in particular the
statistics of the transverse derivative of the streamwise turbulent velocity component 
$\partial u_x^{\prime}/\partial y$ is a sensitive measure for detecting deviations from local
isotropy in homogeneous or nearly homogeneous shear flows.\cite{Warhaft2002,Schumacher2003,Biferale2005}
In a shear flow with a mean shear rate $S>0$, one expects a positive value for derivative skewness 
and other higher odd order moments which are defined as 
\begin{equation}
M_{2n+1}(\partial u_x^{\prime}/\partial y)=\frac{\langle(\partial u_x^{\prime}/\partial y)^{2n+1}\rangle}
              {\langle(\partial u_x^{\prime}/\partial y)^{2}\rangle^{n+1/2}}\,.
\end{equation}
The derivative moments would be exactly zero in a perfectly isotropic flow. 
Their non-zero magnitudes indicate that velocity gradient fluctuations of the streamwise component
along the direction of the outer shear gradient are more
probable than the ones in the opposite direction. 
It can be expected that the asymmetry in the angular distribution, which we discussed above, 
will have an impact
on the statistics of exactly these gradient fluctuations. Figure~\ref{fig9} reports our findings for
the PDF of the transverse derivative, which has been normalized by its root mean square value for all
cases. We observe in both figures a depletion of the left hand tail, which stands exactly for the velocity
gradient fluctuations opposite to the direction of the mean shear. The results suggest that the 
preferential orientation fluctuations of the dumbbells at azimuthal angles $\varphi>0$ go in line with a
depletion of the negative tail of the PDF of the transverse derivative. As sketched in Fig.~\ref{fig8},
negative transverse gradients would be amplified by prefential orientations with $\varphi<0$ which correspond to the
dumbbell colored in gray. The findings are consistent with our observations on the $\varphi$-statistics.
They can also be rationalized (but not explained) when considering the equation for the 
Brownian dynamics of the FENE dumbbell \cite{Oettinger1996}
\begin{equation}
\frac{\mbox{d}{\bm R}}{\mbox{d}t}=
{\bm R\cdot\nabla u}-\frac{{\bm R}}{2\tau(1-R^2/L_0^2)}+\sqrt{\frac{R_0^2}{\tau}}{\bm \xi}_{\bm R}\,.
\end{equation}
In the plane shear flow geometry the component $R_x$ along the mean flow direction is of particular 
interest. Since we are interested in stretched dumbbells with $R_x>R_0$ and in $\Wi>1$ we neglect
contributions from the spring force and the noise for a moment. With the Reynolds decomposition 
(\ref{Reynolds}) we get
\begin{eqnarray}
\frac{\mbox{d}R_x}{\mbox{d}t}&\simeq& \left( S+ \frac{\partial u_x^{\prime}}{\partial y} \right) R_y
+\left(\frac{\partial u_x^{\prime}}{\partial x} \right) R_x + ...\,,\label{r1b}\\
\frac{\mbox{d}R_y}{\mbox{d}t}&\simeq& \left( \frac{\partial u_y^{\prime}}{\partial y} \right) R_y+
\left( \frac{\partial u_y^{\prime}}{\partial x} \right) R_x+
...\label{r2b}
\end{eqnarray}
The important term is the first term on the r.h.s. of (\ref{r1b}). The other three contributions will
behave as noise terms. Fluctuating gradients $\partial u_x^{\prime}/\partial y$ along $S{\bm e}_y$ lead
to a more rapid growth of $R_x$ (for an angle $\varphi>0$) and a prefered alignment with the mean flow. 
This causes
a more rapid decrease of $R_y$ and consequently of $R_x$ via (\ref{r1b}). 
The dumbell can be kicked afterwards again to 
larger $\varphi$ values and transfers momentum to the flow which corresponds exactly to a local patch of 
$\partial u_x^{\prime}/\partial y>0$ (see also Fig. (\ref{fig8})). Then $R_y$ grows and this whole cycle starts
anew. Small scale gradients with the opposite sign diminish the total shear in the surrounding of the 
dumbbell and cause a less efficient stretching and cycle. Clearly, this picture omits some important
features such as the tumbling of the dumbbells.
 
The depletion of gradient fluctuations goes in line with experimental observations by Liberzon 
{\it et al.} \cite{Liberzon2005,Liberzon2006} The authors found e.g. that the 
enstrophy production became 
anisotropic when polymers are added to the fluid. This quantity is directly related to transverse 
gradient components discussed here.   

\subsection{Invariants of the velocity gradient tensor and dumbbell extension} 
The efficient stretching of the dumbbells is connected to particular local flow 
topologies. They are related to the three eigenvalues $\lambda_i$ of the velocity gradient tensor or the 
corresponding three velocity gradient tensor invariants, which are denoted as $I_1$, $I_2$, and $I_3$. 
The eigenvalues of the velocity gradient tensor 
$\partial u_i^{\prime}/\partial x_j$ result as zeros of the following third-order characteristic 
polynomial\cite{Davidson} 
\begin{equation}
\lambda^3- I_1\lambda^2 + I_2\lambda -I_3 =0\,.
\label{polynomial}
\end{equation}
For an incompressible flow \footnote{We will use the definitions as given in Ref.~\cite{Davidson} (p. 264) with $I_1=P$, $I_2=Q$,
and $I_3=R$},
\begin{eqnarray}
I_1&=&\lambda_1 + \lambda_2 +\lambda_3=\mbox{Tr}\left(\frac{\partial u_i^{\prime}}{\partial x_j}\right)=0 \,,\nonumber\\
I_2&=&\lambda_1\lambda_2+\lambda_2\lambda_3+\lambda_3\lambda_1=-\frac{1}{2}\,
\frac{\partial u_i^{\prime}}{\partial x_j}\,\frac{\partial u_j^{\prime}}{\partial x_i} \,,\nonumber\\
I_3&=&\lambda_1 \lambda_2 \lambda_3=\mbox{det}\left(\frac{\partial u_i^{\prime}}{\partial x_j}\right)=
\frac{1}{3}\,\frac{\partial u_i^{\prime}}{\partial x_j}\,\frac{\partial u_j^{\prime}}{\partial x_k}\,
             \frac{\partial u_k^{\prime}}{\partial x_i} \,.
\label{pqrdef}
\end{eqnarray}
The remaining coefficients of (\ref{polynomial}) are therefore $I_2$ and $I_3$, which span the $I_3-I_2$
parameter plane. The scatter 
plots for turbulent flows result in a typical skewed teardrop shape. With our definitions given above the
following crude classification scheme can be given. For $I_2>0, I_3>0$ vortex stretching is present
corresponding to $\lambda_1=a, \lambda_{2,3}=-a\pm\mbox{i}b$ (first quadrant); for $I_2>0, I_3<0$ 
vortex compression is
present corresponding to $\lambda_1=-a, \lambda_{2,3}=a\pm\mbox{i}b$ (second quadrant). The cases 
$I_3<0$ are 
associated with bi-axial strain for $I_2<0$ (third quadrant) corresponding to
$\lambda_1=a, \lambda_{2}=b, \lambda_3=-(a+b)$ and with uniaxial strain at $I_2>0$ (fourth quadrant)
corresponding to
$\lambda_1=a, \lambda_{2}=-b, \lambda_3=-(a-b)$. Constants $a$ and $b$ are larger than zero in all cases.      
Figure~\ref{fig10} relates the extension of the dumbbells to the corresponding local velocity gradients
in the $I_3-I_2$ plane (and consequently to the existing local flow topology). The invariants of the
velocity gradient were evaluated in the center of mass of each dumbbell. The typical teardrop
shape for the turbulence data in the parameter plane is detected. 

Our findings can be summarized as follows. Strongly stretched dumbbells go in line with the largest
excursions of the gradients in the $I_3-I_2$ plane. The longest dumbbells are found preferentially in
regions where vortex stretching or bi-axial strain dominate the local flow topology. The preferential
stretching by bi-axial strain was discussed already for the passive advection of FENE dumbbells in a 
minimal flow unit.\cite{Terrapon2004} It corresponds to the scenario that different parts of the 
dumbbell get pulled by counterstreaming streamwise streaks. The preferential extension close to vortex
stretching means that the polymers are pulled around streamwise vortices. This point was outlined in 
Ref.~\cite{Dubief2004} on the basis of an analysis of the energetics of viscoelastic turbulence. Here,
we find both in a common description based on the analysis of the full velocity gradient tensor,
i.e. the symmetric strain tensor plus the anti-symmetric vorticity tensor. We do also observe that 
the area of the teardrop shape shrinks with increasing Stokes number. This indicates that the small-scale
velocity gradients are supressed in magnitude, which goes in line with more limited excursions across the 
$I_3-I_2$ plane and a relaminarization of the turbulence. Again, this goes in line with very recent 
experimental observations by Liberzon {\it et al.}\cite{Liberzon2006}
   
\section{Summary and discussion}
The presented numerical studies aimed at connecting a macroscopic description for the Newtonian
turbulent shear flow to the mesoscopic description of an ensemble of FENE dumbbells which are advected 
in such flow. The momentum transfer of the dumbbells with the fluid is implemented by an additional 
volume forcing in the Navier-Stokes equations. In numerical terms,
pseudospectral simulations for the solvent are coupled to a system of stochastic nonlinear ordinary 
equations in order to model a viscoelastic fluid.

For the accessible parameters we found slight modifications of the macroscopic flow
structures and mean statistical properties only. This was demonstrated for the global energy 
balance and the mean components of the Reynolds stress tensor. We conclude that dumbbell inertia effects
are present, but remain subleading in comparison to the elastic properties. For the present viscoelastic
flow a drag reduction of up to 20\% is achieved. The microscopic properties of turbulence
were found to be more sensitive with respect to the Weissenberg number. The statistics of the azimuthal
angle $\varphi$ is consistent with former findings for elastic Hookean dumbbells. \cite{Schumacher2006} A growing
number of dumbbells becomes increasingly aligned with the mean flow direction.   
The feedback of the FENE dumbbells on the small-scale properties of turbulence is demonstrated for two 
gradient measures, the PDF of the transverse derivative of the turbulent streamwise velocity component
$\partial u_x^{\prime}/\partial y$ and the diminished scattering of the velocity gradient invariants 
ampiltudes in the $I_3-I_2$ plane with increasing $\Wi$.
The asymmetry of the PDF $p(\partial u_x^{\prime}/\partial y)$ is found to increase with increasing 
$\Wi$. Furthermore, we determined that strongly
stretched dumbbells can be found close to vortex stretching or biaxial strain topologies of the advecting
shear flow.       

The present study should be considered as a first step for such class of hybrid models. 
One difference to the situation in a dilute polymer solution is the relatively large Stokes number
that had to be taken. Our dispersed dumbbells behave in parts like deformable particles rather than 
polymer chains. Frequently, heavier quasi-particles are used for the study of turbulence in 
particle-ladden flows.\cite{Bosse2006} Extensions 
of our investigations will have to go into two directions. Firstly, it is desirable that larger spectral 
resolutions, like the ones in Ref.~\cite{Schumacher2006}, are achieved. This will require a fully parallel
implementation of the current numerical scheme. Larger computational grids and higher Reynolds numbers
will give us the opportunity to
decrease the ratio $R_0/\etak$ and to increase $L_0/R_0$ to more realistic values. 
Secondly, eq. (\ref{ratio}) implies the efforts that have to be taken in order to approach the situation
in a polymer solution. Decreasing values of $R_0$ and $\St$ have to be compensated by $\knp$, e.g., a 
reduction 
of both -- $R_0$ and $\St_{\eta}$ -- by an order of magnitude requires an increase of the concentration
(or number density) by a power of 5/2. Once such operating point is reached, the time scale argument
which is thought to be important for the drag reduction effect, can also be studied.\cite{Lumley1969} 
Finally, a recent work by
Vincenzi and co-workers \cite{Vincenzi2006} provides an interesting ansatz for modelling the polymer
dynamics. The authors studied a
conformation-dependent Stokes drag coefficient that caused a significant dynamical slow-down of the
coil-stretch transition in steady elongational and random flows. The test of these ideas in turbulent
shear flows is still to be done.

\acknowledgements
This work was supported by the Deutsche Forschungsgemeinschaft (DFG) and the Deutscher
Akademischer Austauschdienst (DAAD) within the German-French PROCOPE program. We thank
for computing ressources on the JUMP supercomputer at the John von Neumann Institute for Computing,
J\"ulich (Germany). Further computations have been conducted at the MARC cluster (Marburg)
and the MaPaCC cluster (Ilmenau). Fruitful discussions with F. de Lillo, B. Eckhardt, and D. Vincenzi
are acknowledged.

\appendix
\section{Semi-implicit integration scheme for dumbbells}
The FENE dumbbells consist of two beads at positions ${\bm x}_1(t)$ and ${\bm x}_2(t)$ which are
connected by a nonlinear elastic spring. The velocities of the advecting flow at both beads are
denoted by ${\bm u}_1$ and ${\bm u}_2$, respectively. Note that these velocities coincide with
$\dot{\bm x}_1$ and $\dot{\bm x}_2$, respectively, for $\St=0$ only. Since the beads are usually found 
between mesh vertices, the values for ${\bm u}_1$ and ${\bm u}_2$ have to be determined by trilinear
interpolation from the known velocity vectors at the neighboring grid sites. The dynamical
equations for the dumbbells are set up in relative and center-of-mass coordinates. The relative 
coordinate (or separation) vector of the dumbbell is given by
\begin{equation}
{\bm R}(t)={\bm x}_2(t)-{\bm x}_1(t)\,.
\end{equation}
The center-of-mass coordinate vector is given by 
\begin{equation}
{\bm r}(t)=\frac{1}{2}({\bm x}_1(t)+{\bm x}_2(t))\,.
\end{equation}
The velocities which are assigned with the relative and center-of mass coordinates are denoted as
${\bm V}$ and ${\bm v}$, respectively.     
The Newtonian equations for the dynamics of the FENE dumbbells in dimensionless form,  which
follow then from (\ref{r1})-(\ref{r4}) with the
definitions (\ref{Stokes}) and (\ref{Weiss}), are given by
\begin{eqnarray}
\frac{\mbox{d}\tilde{\bm r}}{\mbox{d}t}&=&\tilde{\bm v} \label{d1}\,,\\
\frac{\mbox{d}\tilde{\bm v}}{\mbox{d}t}&=&\frac{1}{\St}\left[-\tilde{\bm v}+
\frac{1}{2}(\tilde{\bm u}_1+\tilde{\bm u}_2)+
\frac{R_0}{2L\sqrt{\Wi}}\tilde{\bm \xi}_{\bm r}\right] \label{d2}\,,\\
\frac{\mbox{d}\tilde{\bm R}}{\mbox{d}t}&=&\tilde{\bm V} \label{d3}\,,\\
\frac{\mbox{d}\tilde{\bm V}}{\mbox{d}t}&=&\frac{1}{\St}
\left[-\tilde{\bm V}+(\tilde{\bm u}_2-\tilde{\bm u}_1)-\frac{\tilde{\bm R}}
{2\Wi\left(1-\tilde{R}^2 L^2/L_0^2\right)}+
            \frac{R_0}{\sqrt{\Wi}\,L}\tilde{\bm \xi}_{\bm R}\right]\,.\label{d4}
\end{eqnarray}
For the following, we omit the tilde symbol for the dimensionless quantities.
The predictor values of the center-of-mass vector ${\bm r}$ and the distance
vector ${\bm R}$ are calculated by an explicit Euler step whereas the corresponding velocities are
treated by an implicit Euler step, giving
\begin{eqnarray}
{\bm r}^*&=&{\bm r}^l+\Delta t{\bm v}^l \label{p1}\,,\\
{\bm v}^*&=&\frac{1}{\St+\Delta t}\left[\St\,{\bm v}^l+
\frac{1}{2}({\bm u}_1^l+{\bm u}_2^l)\Delta t+
\frac{R_0}{2L\sqrt{\Wi}}\Delta {\bm w}^l\right]\label{p2}\,,\\
{\bm R}^*&=&{\bm R}^l+\Delta t{\bm V}^l \label{p3}\,,\\
{\bm V}^*&=&\frac{1}{\St+\Delta t}\left[\St\,{\bm V}^l+
({\bm u}_2^l-{\bm u}_1^l)\Delta t-
\frac{{\bm R}^l}{2 \Wi \left(1-(R^l)^2 L^2/L_0^2\right)}\, \Delta t+
\frac{R_0}{L\sqrt{\Wi}}\Delta {\bm W}^l\right]\,.
\label{p4}
\end{eqnarray}{
The corrector step for the center-of-mass and distance vectors is given as
\begin{eqnarray}
{\bm r}^{l+1}&=&{\bm r}^l + \frac{1}{2} ({\bm v}^*+{\bm v}^l) \Delta t\label{c1}\\
{\bm v}^{l+1}&=&\frac{1}{\St+\Delta t}\,\Bigg[\frac{1}{2} \left(\St\,{\bm v}^*
 + \frac{1}{2}({\bm u}^*_1+ {\bm u}^*_2) \Delta t+\St\,{\bm v}^l +
   \frac{1}{2}({\bm u}^l_1+ {\bm u}^l_2) \Delta t\right)+\nonumber\\
& &\frac{R_0}{2 L \sqrt{\Wi}}\Delta {\bm W}^l\Bigg]\label{c2}\\
{\bm R}^{l+1}&=&{\bm R}^l + \frac{1}{2} ({\bm V}^l+{\bm V}^{l+1}) \Delta t\label{c3}\\
{\bm V}^{l+1}&=&\frac{1}{\St+\Delta t}\,\Bigg[\frac{1}{2} \Bigg(\St\,{\bm V}^*
 + ({\bm u}^*_2- {\bm u}^*_1) \Delta t+ \St\,{\bm V}^l +
   ({\bm u}^l_2- {\bm u}^l_1) \Delta t-\nonumber\\
& & \frac{{\bm R}^{l+1}}{2\Wi\left(1-({\bm R}^{l+1})^2 L^2/L_0^2\right)}\,\Delta t-
  \frac{{\bm R}^{l}}{2\Wi\left(1-({\bm R}^{l})^2 L^2/L_0^2\right)}\,\Delta t\Bigg)
  +\nonumber\\
& & \frac{R_0}{L\sqrt{\Wi}}\Delta {\bm W}^l]\,.\label{c4}
\end{eqnarray}
Note that the corrector step for the distance vector is semi-implicit in the velocity in order 
to avoid stiffness of the equation system at small Stokes numbers. The corrector step 
for the distance velocity 
${\bm V}$ has to be semi-implicit in the separation vector ${\bm R}$ due to the finite extensibility of
the dumbbells.\cite{Oettinger1996} When inserting (\ref{c4}) into (\ref{c3}) one gets
\begin{equation}
\left(1+\frac{(\Delta t)^2}{8\Wi\,(\St+\Delta t)
\left(1-({\bm R}^{l+1})^2 L^2/L_0^2\right)}
\right){\bm R}^{l+1}={\bm A}\,,
\label{poly1}
\end{equation}
where the abbrevation ${\bm A}$ contains terms only which are known. By taking the norm of
(\ref{poly1}) one ends up with a cubic polynomial for $R^{l+1}$. The formula for the 
``casus irreducibilis" of three real solutions of the polynomial goes back to F. Vi\`{e}te 
\cite{Vieta} and yields directly the unique solution for $R=|{\bm R}|$ between 0 and $L_0$.
From (\ref{poly1}) follows now 
\begin{equation}
{\bm R}^{l+1}=R^{l+1}\frac{\bm A}{A}\,.
\label{poly2}
\end{equation}
This value is inserted into (\ref{c4}) which completes the corrector step.

\begin{figure}
\centerline{\includegraphics[scale=0.8]{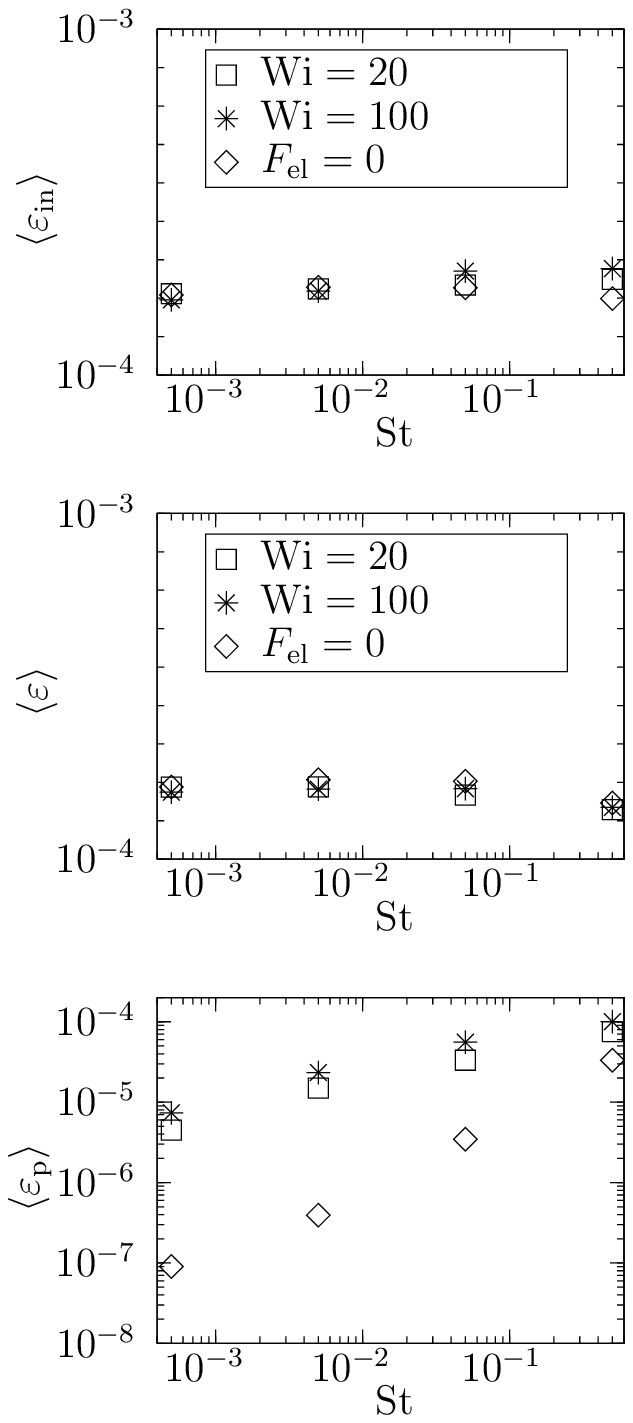}}
\caption{Mean dissipation and injection rates as a function of the Stokes
and Weissenberg numbers.
Upper picture: mean energy injection rate due to shear flow forcing
$\langle\epsin\rangle$. Mid panel: mean energy dissipation rate $\langle\varepsilon\rangle$.
Lower panel: mean dissipation rate which
arises from the coupling to the dumbbell ensemble $\langle\epsp\rangle$. The Reynolds
number is $\Rey=800$. The case with $F_\mathrm{el}=0$ is for the case without spring force and stands for a
shear flow with dispersed inertial particles.}
\label{fig1}
\end{figure}
\begin{figure}
\centerline{\includegraphics{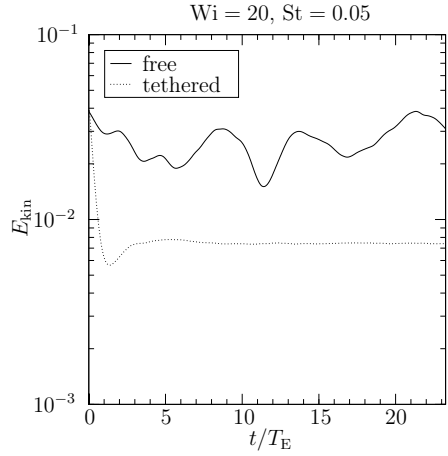}}
\caption{Comparison of the kinetic energy for two cases at $\Wi=20$ and $\St=0.05$:
for the tethered case one bead of each dumbbell is fixed at a grid site while the second bead
can fluctuate. The freely draining case is the usual situation which allows the free motion
of the dumbbells through the turbulent flow volume. The Reynolds number is $\Rey=400$.}
\label{fig2}
\end{figure}
\begin{figure}
\centerline{\includegraphics{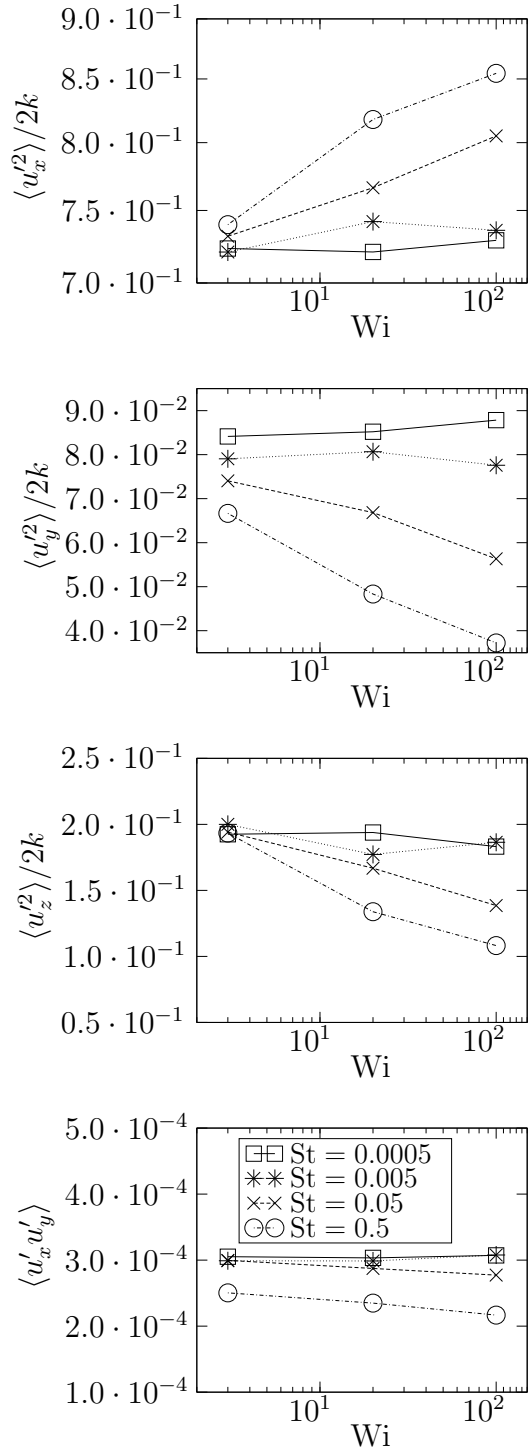}}
\caption{Reynolds stresses $\langle u_i^{\prime} u_j^{\prime}\rangle$
normalized by the turbulent kinetic energy
$k=\langle (u_i^{\prime})^2\rangle/2$ as a function of the Weissenberg and Stokes numbers.
From top to bottom: streamwise fluctuations, fluctuations in shear direction, spanwise
fluctuations, and shear stress. The Reynolds number is $\Rey=800$.}
\label{fig3}
\end{figure}
\begin{figure}
\centerline{\includegraphics{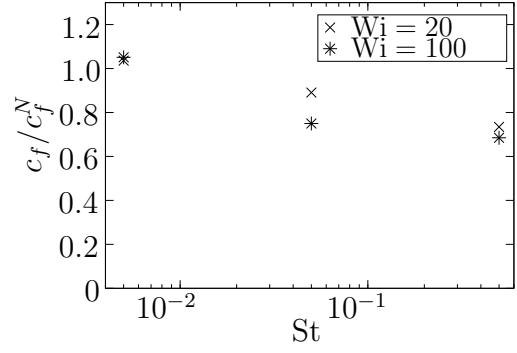}}
\caption{Ratio of friction factors as a function of $\St$ for the largest $\Wi$.
The friction factor for the fluid with the dispersed dumbbells is $c_f$
(cf. Eq.~(\ref{friction})). The quantity $c_f^N$ is the friction factor of the
Newtonian fluid.}
\label{fig4}
\end{figure}
\begin{figure}
\centerline{\includegraphics[scale=0.3]{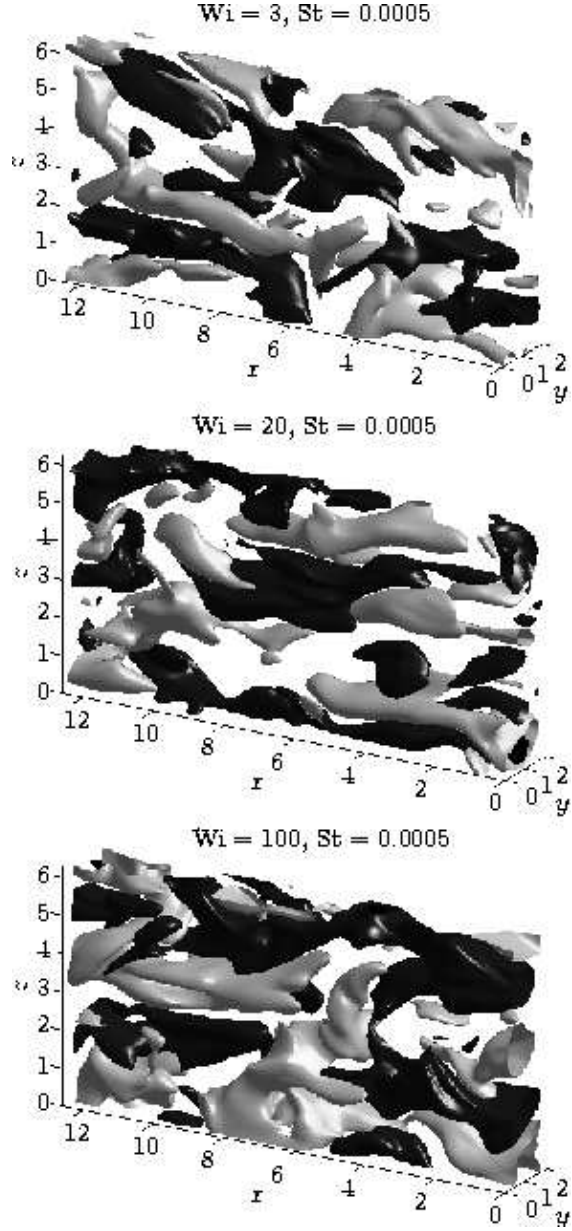}}
\caption{Isosurface plot of the fluctuations of the streamwise turbulent velocity component
$u_x^{\prime}$. The snapshots are for $\Rey=800$ and $\St=0.0005$. The isolevels are for $\pm 0.04$ in
each case.}
\label{fig5}
\end{figure}
\begin{figure}
\centerline{\includegraphics{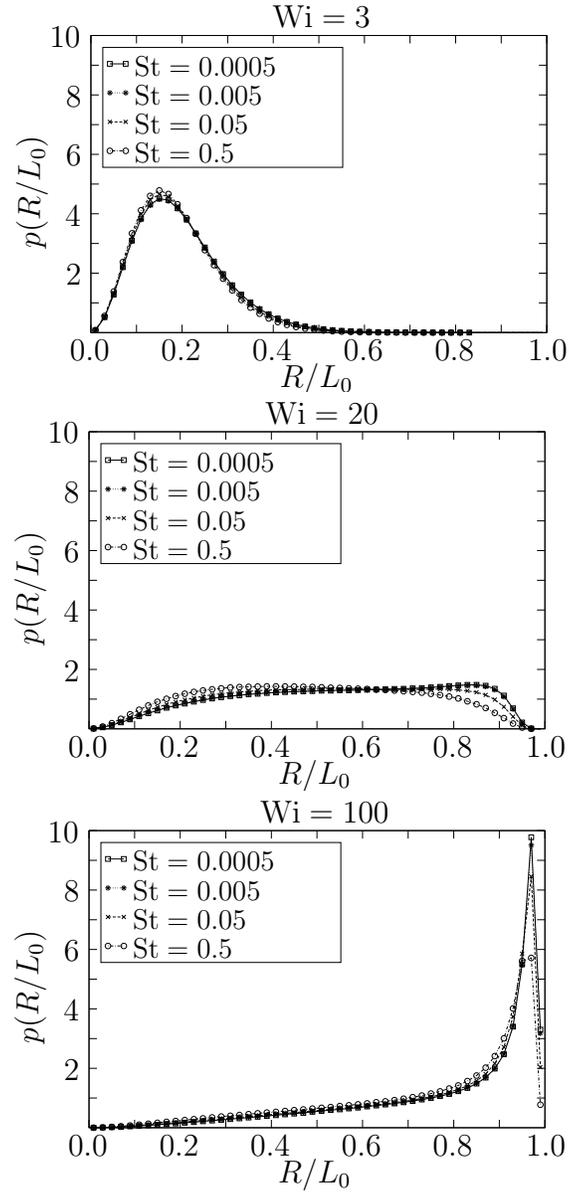}}
\caption{Probability density function (PDF) of the extension $R$ normalized by the contour
length $L_0$. Three different Weissenberg numbers are shown. The Stokes numbers of the data
are indicated in the legend. Data are for $\Rey=800$.}
\label{fig6}
\end{figure}
\begin{figure}
\centerline{\includegraphics{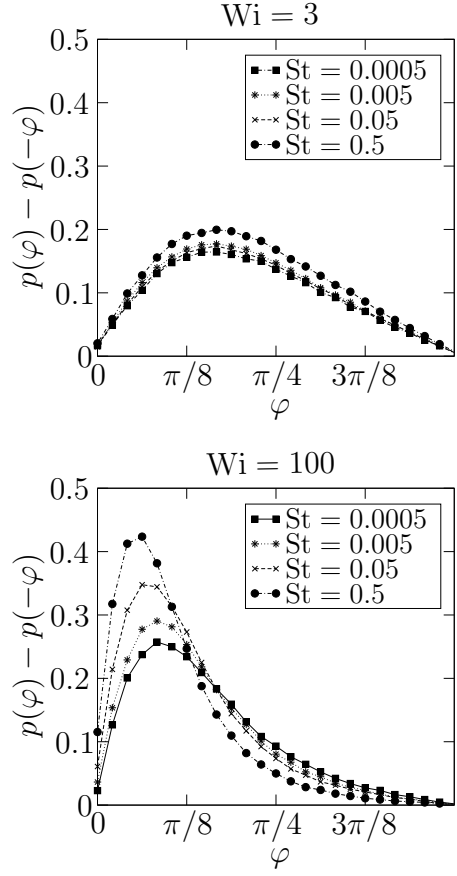}}
\caption{Asymmetry of the probability density function (PDF) of the azimuthal angle
$\varphi$. It is defined as $A(\varphi)=p(\varphi)-p(-\varphi)$. The upper panel shows the data for
$\Wi=3$ and four different Stokes numbers. The lower panel shows the data for
$\Wi=100$ and four different Stokes numbers. The analysis is for $\Rey=800$.}
\label{fig7}
\end{figure}
\begin{figure}
\centerline{\includegraphics{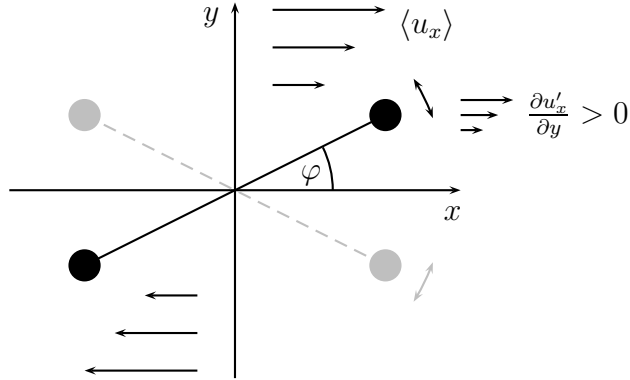}}
\caption{Sketch of the orientation of a dumbbell in the turbulent shear flow.
The mean turbulent flow profile is indicated.
The dark-colored dumbbell stands for the preferentially oriented one while the
gray-colored orientation is less probable. This orientation asymmetry leads to the asymmetry 
in the angular distribution as given
in Fig.~\ref{fig7}.}
\label{fig8}
\end{figure}
\begin{figure}
\centerline{\includegraphics{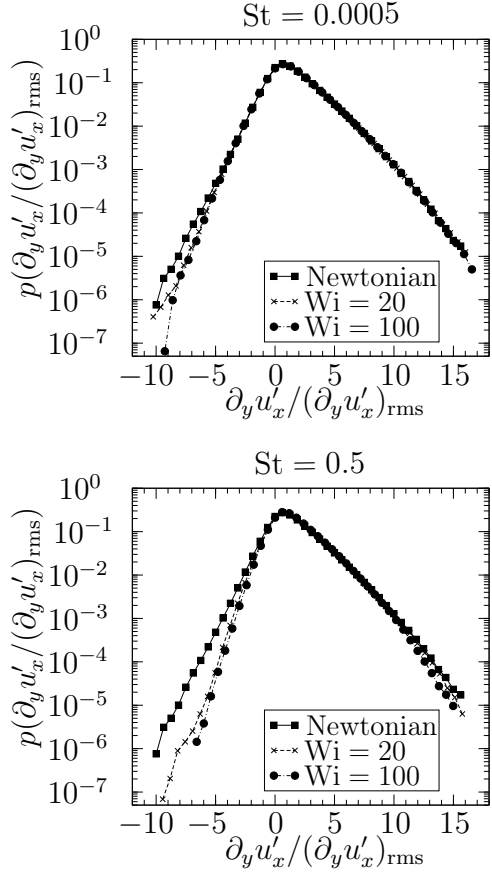}}
\caption{Probability density function (PDF) of the transverse velocity gradient of the streamwise
turbulent fluctuations, $\partial u_x^{\prime}/\partial y$. The Newtonian case is compared with the
two larger values of the Weissenberg number at $\St=5\times 10^{-4}$ and 0.5, respectively. The data are
for $\Rey=800$.}
\label{fig9}
\end{figure}
\begin{figure}
\centerline{\includegraphics[scale=0.7]{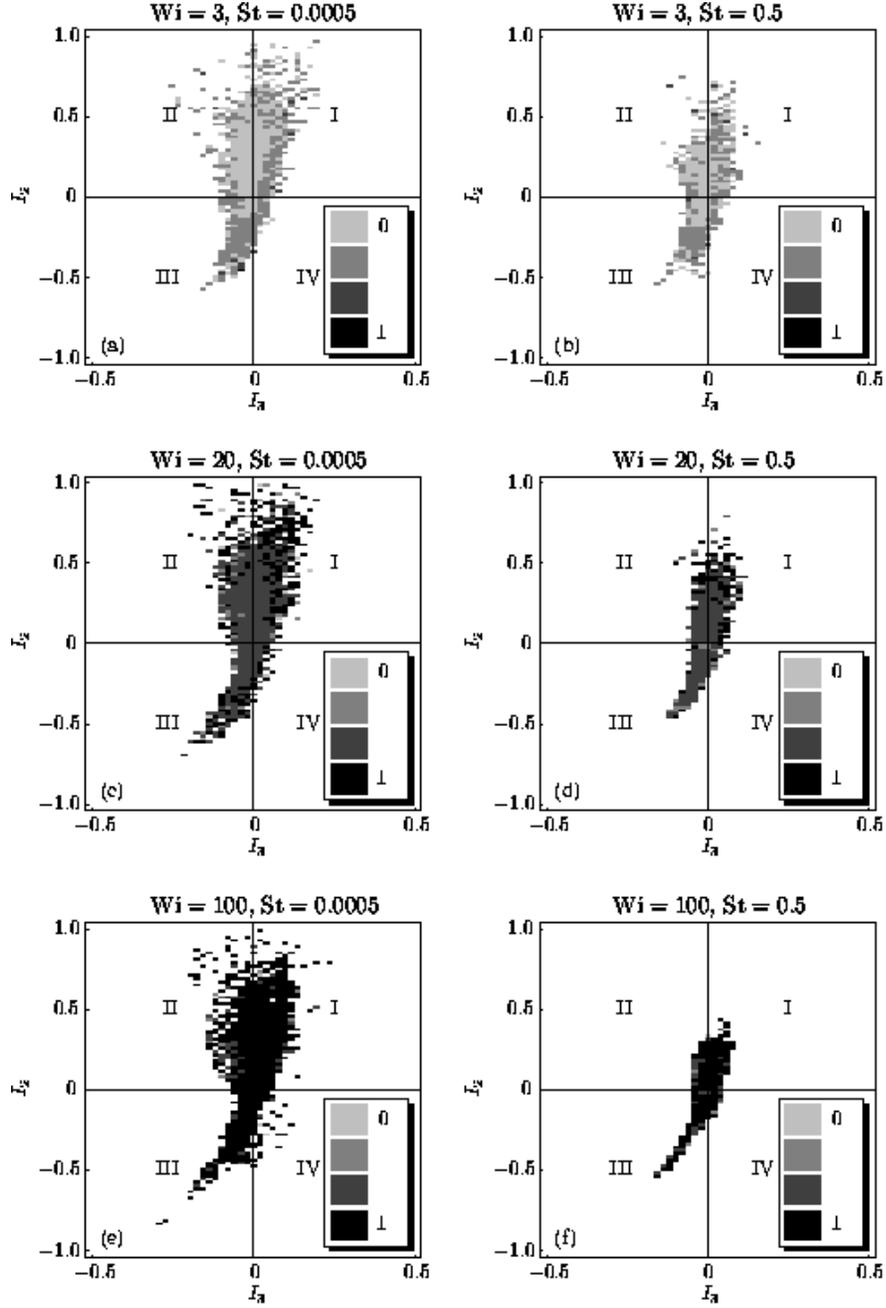}}
\caption{Relation between the extension of the dumbbells and the local velocity gradient
at the center of mass of the dumbbells. The local flow topology that is related to
the velocity gradient is quantified by the second and third invariants $I_2$ and $I_3$ (see eqns. (\ref{pqrdef})
for the definition). Quadrant I stands for vortex stretching, II for vortex compression, III for bi-axial
strain, and IV for uniaxial strain, respectively.  The gray color coding of the bins for $0< R/L_0<0.25$,
$0.25\le R/L_0<0.5$, $0.5\le R/L_0<0.75$, $0.75\le R/L_0\leq1$ is indicated by the legend for
each figure. Data are for $\Rey=800$.}
\label{fig10}
\end{figure}

\end{document}